\documentclass[usenatbib]{mn2e}

\usepackage{rotating}
\usepackage{graphics}
\usepackage{aas_macros}

\newcommand{\teff}{$T_{\rm eff}$}
\newcommand{\dnlte}{$\rm \Delta_{NLTE}$}
\newcommand{\eexc}{$E_{\rm exc}$}

\def\vt{$\xi_{\rm t}$}
\def\kms{$\rm km~s^{-1}$}
\def\ione{\,{\sc i}}
\def\ii{\,{\sc ii}}

\newcommand{\eps}{\log\varepsilon}
\title[Li-rich VMP giant]{The Pristine survey – XXII. A serendipitous discovery of an extremely Li-rich very metal-poor giant and a new method of $^6$Li/$^7$Li isotope measurement\thanks{Based on observations made with the Subaru Telescope.}}

\author[Sitnova et al.]{
\parbox{\textwidth}{T. M. Sitnova\thanks{E-mail:sitamih@gmail.com}, T. Matsuno$^{1}$, Z. Yuan$^{2}$, N. F. Martin$^{2}$, P. Banerjee$^{3}$, F. Sestito$^{4}$, K. A. Venn$^{4}$, J. I. Gonz\'alez Hern\'andez$^{5,6}$}\\
\\
$^{1}$Kapteyn Astronomical Institute, University of Groningen, Landleven 12, 9747 AD Groningen, The Netherlands\\
$^{2}$Universit\'e de Strasbourg, CNRS, Observatoire Astronomique de Strasbourg, UMR 7550, F-67000 Strasbourg, France \\
$^{3}$Department of Physics, Indian Institute of Technology Palakkad, Kerala 678558, India\\
$^{4}$Department of Physics and Astronomy, University of Victoria, PO Box 3055, STN CSC, Victoria BC V8W 3P6, Canada\\
$^{5}$Instituto de Astrof{\'\i}sica de Canarias, E-38205 La Laguna, Tenerife, Spain \\
$^{6}$Universidad de La Laguna, Dept. Astrof{\'\i}sica, E-38206 La Laguna, Tenerife, Spain
}
\date{Accepted XXX. Received YYY; in original form ZZZ}
\pubyear{}
\begin{document}
\label{firstpage}
\pagerange{\pageref{firstpage}--\pageref{lastpage}}
\maketitle
\begin{abstract}
We report the serendipitous discovery of a very metal-poor (VMP) Li-rich giant star (\teff\ = 4690$\pm$80~K, log~g = 1.34$\pm$0.13, [Fe/H] = $-2.43\pm$0.07). We analyse the Li\ione\ 6103 and 6707 \AA\ lines accounting for departures from local thermodynamic equilibrium (NLTE) and correcting for 3D effects using literature data, which yields a lithium abundance $\eps_{Li} = 3.42\pm0.07$. Comparing lithium abundances from the two lines, in 1D~NLTE we measure the isotope ratio $^6$Li/$^7$Li =  1.64$^{+1.49}_{-1.08}$ \%. When correcting for 3D effects, we detect the fragile $^6$Li isotope at $2$-sigma level and the ratio $^6$Li/$^7$Li = 5.65$^{+5.05}_{-2.51}$ \%. To our knowledge, this is the first $^6$Li/$^7$Li measurement in an extremely Li-rich VMP star. The Cameron-Fowler mechanism, which is proposed to produce Li-rich stars, does not imply $^6$Li production and is therefore inconsistent with our measurement when applying 3D corrections. We also derive NLTE abundances for 16 elements, most of which show similar abundances to those found in VMP stars. Sodium is an exception: [Na/Fe]$_{\rm NLTE 1D}$ = 0.07 $\pm 0.03$, which is 0.5\,dex higher than what is typical for VMP stars. This star joins the sample of rare Li-rich VMP stars, and we offer a novel way to constrain the source of lithium in such stars through isotope ratio measurements.

\end{abstract}


\section{Introduction}
\label{sec:intro}

Lithium is an element with a complex astrophysical origin and chemical evolution. 
It is one of the primordial elements produced in the Big Bang and can also be produced through further processes, such as the spallation process and stellar nucleosynthesis.  
A number of processes have been suggested as sources of lithium, but so far its exact origins and production mechanisms remain unclear \citep[see, for example,][]{2012A&A...542A..67P,2021Msngr.185...18M,2021A&A...653A..72R}.

Lithium has two stable isotopes: $^7$Li and the less abundant and more fragile $^6$Li. 
The two isotopes have different origins: 
$^7$Li is produced in the Big Bang nucleosynthesis, inside stars, and by cosmic rays via spallation, while $^6$Li can only be produced by cosmic rays via spallation \citep{2012A&A...542A..67P}. 
As a result, the ratio between $^7$Li and $^6$Li varies among different astrophysical sites. 
For example, in Solar system meteorites, the ratio is measured to be $^6$Li/$^7$Li = 8.11~\% \citep{2003LPI....34.1931M}, while \citet{2017A&A...604A..44M} found the same value in a metal-rich magnetically active giant star, and \citet{1997AA...328..695R} found $^6$Li/$^7$Li = 3 \%\ and 8 \%\ in solar spots and in active late type dwarf stars, respectively. The most up to date NLTE calculations of the Li\ione\ 6707 \AA\ line in 3D model atmospheres indicate the absence of $^6$Li in the sun \citep{2018A&A...612A..44S} and solar type stars \citep{2018A&A...618A..16H}.

In old, very metal-poor (VMP, [Fe/H]\footnote{We use a standard designation, [X/Y] = $\log($N$_{\rm X}$/N$_{\rm Y}$)$_{*} - \log($N$_{\rm X}$/N$_{\rm Y}$)$_{\odot}$, where N$_{\rm X}$ and N$_{\rm Y}$ are total number densities of element X and Y, respectively.} $< -2$) stars with normal lithium abundance, the presence of  the $^6$Li isotope is unlikely. Some of the early studies \citep{1993ApJ...408..262S,1998ApJ...506..405S,2006ApJ...644..229A} report on $^6$Li detections in unevolved VMP stars, while others \citep[for example,][]{2009A&A...504..213G} conclude that the detection of $^6$Li cannot be safely claimed. The above studies employ classic 1D model atmospheres, which neglect convection.
\citet{2007A&A...473L..37C} and \citet{2012MSAIS..22..152S} showed that the convective asymmetry generates an excess absorption in the red wing of the resonance line that mimics the presence of $^6$Li and the measurements of ratios for unevolved stars should be considered as upper limits. 
\citet{2013A&A...554A..96L}, \citet{2019A&A...628A.111G}, and \citet{2022MNRAS.509.1521W} account for deviations from local thermodynamic equilibrium (LTE) and hydrodynamic (3D) effects and confirm non-detections of $^6$Li in these stars.

The majority of unevolved VMP stars have normal lithium abundance $\eps_{Li}$\footnote{Here, $\eps$ = log N$_{\rm El}$/N$_{\rm H}$, where N$_{\rm El}$ and N$_{\rm H}$ are number densities of a given chemical element and hydrogen, respectively.} = 2.25, known as the lithium plateau or the "Spite plateau" \citep{1982A&A...115..357S}. This value is treated as an upper boundary for lithium abundances in VMP unevolved stars, which decreases with stellar evolution \citep[see, for example,][]{2009A&A...503..545L}.
However, large spectroscopic surveys uncovered a number of very metal-poor stars with lithium abundances exceeding the Spite plateau.
For example, \citet{2018ApJ...852L..31L} found 12 VMP stars, including subgiants, with $\eps_{Li}$  up to 4.53. Nine of them show sodium enhancement, while other measured elements (carbon, magnesium, barium) show values close to those measured in normal stars with similar metallicity. 
\citet{2019A&A...623A..55M,2021A&A...652A.139M} discovered two Li-rich giant stars in $\omega$ Cen. The most Li-rich of these two stars is strongly enhanced in sodium with [Na/Fe]$_{\rm NLTE}$ = 1.01, while another one has [Na/Fe]$_{\rm NLTE}$ = 0.14 in line with other $\omega$ Cen stars.
The most lithium-rich star known to date \citep{2022arXiv220902184K} has  $\eps_{Li}$ = 5.62, [Fe/H] = --2.43 and also shows sodium enhancement with [Na/Fe] = 1.10.
\citet{2012A&A...539A.157M} found a Li- and Na-rich star in the globular cluster M4 and suggested that lithium is produced in parallel to sodium.
However, it is worth noting that stars with high sodium abundances are common and only a small fraction of them are enriched in lithium.

The mechanism of Li-enhancement in unevolved stars is unclear, while in stars at advanced evolutionary stages, $^7$Li can be produced in the Cameron-Fowler mechanism \citep[CF mechanism;][]{1971ApJ...164..111C} at the asymptotic giant branch (AGB) and red giant branch (RGB) stages as proposed by \citet[][]{1971ApJ...164..111C} and \citet{1999ApJ...510..217S}, respectively. In addition to high Li abundance, \citet{1971ApJ...164..111C} predict in some cases high abundances of slow neutron capture (s-process) elements. The most lithium-rich star known to date  \citep{2022arXiv220902184K} can be considered as an example of a star that has undergone the CF mechanism. However, observations show that this is not the only way to produce an excess of lithium in a star \citep{2023arXiv230316124T}. For example, \citet{2017A&A...604A..44M} found a considerable amount of $^6$Li isotope in a magnetically active metal-rich giant, which suggests a different mechanism than CF is responsible for its production. It is important to emphasize that the interpretation of lithium-rich stars varies between those that are metal-poor and those that are metal-rich. This distinction arises because metal-rich stars possess more substantial convective envelopes and experience distinct levels of mixing during their evolution  compared to metal-poor stars. Another reason for  separating the discussion of lithium enhancement in metal-rich and metal-poor stars is the impact of chemical evolution. This is because the contribution of lithium production in novae becomes significant at higher metallicities.

Li-rich stars are ubiquitous, and they are found in different Galactic populations, such as globular clusters \citep[see, for example,][and references therein]{2020A&A...639L...2S}, open clusters \citep{2021A&A...653A..72R}, and dwarf spheroidal galaxies \citep{2012ApJ...752L..16K}.
Over the last few years, continuous efforts have been made on observations of Li-rich stars \citep[see, for example,][]{2016MNRAS.461.3336C,2019ApJS..245...33G,2019MNRAS.484.2000D,2020A&A...639L...2S,2021NatAs...5...86Y,2021MNRAS.505.5340M,2022ApJ...929L..14Y,2022MNRAS.513...71S,2023A&A...671A..61N}. These observations challenge our current theoretical understanding of the origin of lithium and its chemical evolution. To make progress in nucleosynthesis modeling, observational constraints, not only for a given chemical element, but also comprehensive element abundance patterns and isotopic ratios are required. 

In this study, we present the discovery of a Li-rich star and perform a careful stellar parameter and chemical composition determination, including for the $^6$Li/$^7$Li isotopic ratio. 
We present a new method to determine the lithium isotopic ratio. It is based on a comparison of abundances from the resonance line, which is sensitive to the $^6$Li/$^7$Li ratio, and the subordinate line, which is not affected by the $^6$Li/$^7$Li ratio. Our method differs from that used before in the literature since, until now, $^6$Li/$^7$Li ratios were determined by fitting the profile of the resonance line only.
It is worth noting that a similar to our method is used in the literature to derive the barium odd and even isotope ratio: the total barium abundance is determined from the weak subordinate lines and then the isotope ratio is varied until the same abundance is achieved from the saturated resonance lines. The idea was proposed by \citet{1993oee..conf..480M} and applied by \citet{1995A&A...297..686M,2006A&A...456..313M,2019AstL...45..341M}.

We describe the observations and stellar atmosphere parameters in Sec.~\ref{parameters}. The abundance determination method is presented in Sec.~\ref{abund}. The derived chemical element abundances and the $^6$Li/$^7$Li isotopic ratio are presented in Sec.~\ref{results} and Sec.~\ref{67ratio}, respectively. In Sec.~\ref{sec:discussion}, we consider potential scenarios for the high lithium abundance origin in the star of interest. Our conclusions are given in Sec.~\ref{sec:con}. 

\section{Observations and Stellar Parameters}
\label{parameters}

The  star of interest (Gaia DR3 ID = 1918529631627603072, RA = 348.71256241851$^\circ$, DEC = +41.58961513403$^\circ$, $G$ = 13.603 $\pm$ 0.003) is selected from the Pristine-Gaia synthetic  catalog \citep{2023arXiv230801344M} and has a photometric metallicity [Fe/H] = $-2.8^{+0.1}_{-0.2}$. 
It was observed as a backup target of program S22B-094 (PI: Yuan) on the Subaru telescope on September 2022 and was selected as a bright extremely metal-poor ([Fe/H] $< -3$) giant candidate. 

We obtained its high-resolution spectrum with the High Dispersion Spectrograph \citep[HDS,][]{2002PASJ...54..855N}, using the standard StdYd setup, which provides a wavelength coverage of 4000 -- 6800\AA, R = $\Delta \lambda/\lambda$ = 45~000, and signal to noise ratio S/N = 45 around the Li\ione\ lines for our 600-second exposure. The data is reduced using the IRAF\footnote{IRAF is distributed by the National Optical Astronomy Observatory, which is operated by the Association of Universities for Research in Astronomy (AURA) under a cooperative agreement with the National Science Foundation} script hdsql\footnote{\tiny{http://www.subarutelescope.org/Observing/Instruments/HDS/hdsql-e.html}} that includes CCD linearity correction, scattered light subtraction, aperture extraction, flat-fielding, wavelength calibration, and heliocentric velocity correction. 

\begin{figure}
	\includegraphics[width=80mm]{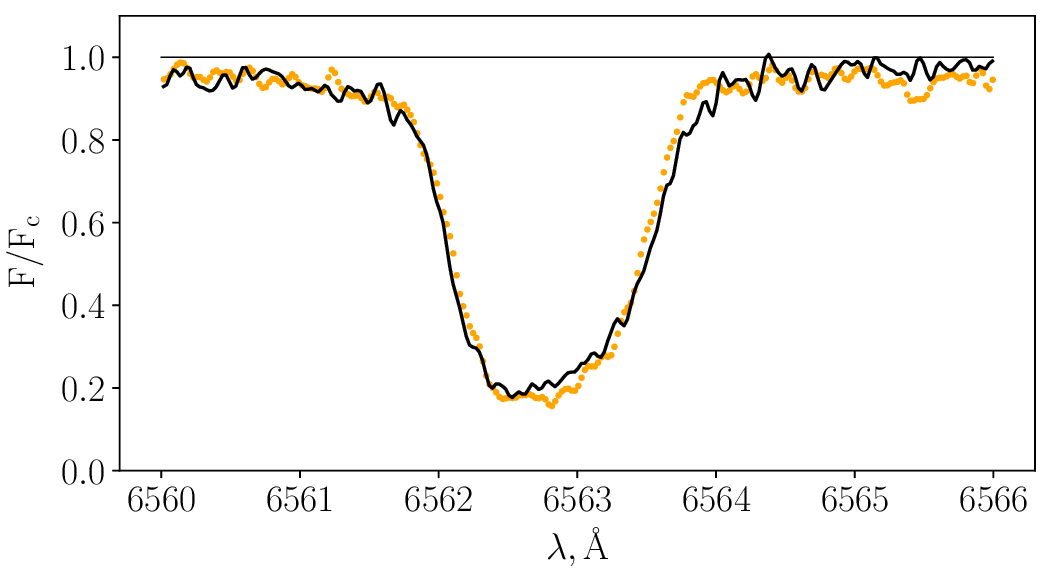}
	\caption{Observed H$_{\alpha}$ line profile in the star of interest (solid curve). For comparison, we show the H$_{\alpha}$ line profile of another star with exactly the same stellar parameters, taken with the same instrument (dots). }
	\label{halpha} 
\end{figure}

The reduced spectrum does not reveal any peculiarities, such as, for example, emission lines. The observed line profiles are narrow and they are not distorted by rapid rotation. From the available spectrum, we do not see signatures of stellar activity. 
\citet{2022ApJ...940...12S} found that half of the Li-rich stars have a strong He\ione\ 10830 \AA\ absorption line, which is an indicator of chromospheric activity and/or mass loss in red giants. While our spectrum does not cover this line, another He\ione\ line at $\lambda$ = 5876 \AA\ is not detected in the available spectrum.
In this regard, a high resolution spectrum covering the He\ione\ 10830 \AA\ and the Ca\ii\ H and K lines would be required  to draw a definite conclusion.  
It is worth noting that the Balmer H$_{\alpha}$ line is slightly asymmetric and the line center is shifted towards the blue, as can be seen in Figure~\ref{halpha} that shows the observed H$_{\alpha}$ profile in the star of interest and another star with nearly the same stellar parameters  \teff/log~g/[Fe/H] = 4650$\pm$80/1.34$\pm$0.24/--2.09$\pm$0.12 (Sitnova et al., in prep.) for comparison. Both spectra were obtained with the same instrument.

From the observed spectrum, we measure a radial velocity V$_{\rm r}$ = -275.1 $\pm$ 0.8 \kms, which is in line with the Gaia measurement V$_{\rm r,\ Gaia}$ = -274.45~$\pm$~1.80 \kms. Gaia DR3 also provides the renormalised unit weight error RUWE = 0.999. Taking into account these data, we assume that this star is a single star.

We calculate an effective temperature \teff\ = 4690 $\pm$ 80~K, a surface gravity log~g = 1.34 $\pm$ 0.13, a metallicity [Fe/H] = $-2.43$ $\pm$ 0.07, and a microturbulent velocity \vt = 1.8 $\pm$ 0.2 \kms.
We determine \teff\ from Gaia $BP-G$, $G-RP$, $BP-RP$ dereddened colors and the calibration of \citet{2021A&A...653A..90M}. The extinction E(B-V) = 0.12 was adopted from \citet{2011ApJ...737..103S} and the colours are corrected according to \citet{2018MNRAS.479L.102C}. Using different colors yields effective temperatures that are consistent within 12~K. The uncertainty on \teff\ is therefore mainly the uncertainty of 80~K on the calibration, as given by \citet{2021A&A...653A..90M}.
For the distance, we calculate $d=~10.1\pm1.5$~kpc using the Gaia parallax, corrected according to \citet{2021A&A...649A...4L}, and following the method of \citet{2015PASP..127..994B}. Given that the uncertainty on the parallax is not large (the ratio between the parallax error and the parallax is 15\%), we determine the distance from the maximum of the distance distribution without invoking a prior.
With the distance, the effective temperature, the bolometric corrections of \citet{2018MNRAS.479L.102C}, and a mass of 0.8 solar masses, we calculate the surface gravity $\log g = 4.44+\log(m/m_{\odot})+0.4(M_{\rm bol} - 4.75) + 4\log($\teff$/5780.0)$, where $m_{\odot}$ is a solar mass and $M_{\rm bol}$ is an absolute bolometric magnitude. The microturbulent velocity is derived from the lines of Fe\ione\ and Fe\ii. The derived stellar atmosphere parameters lead to consistent abundances within 0.02~dex from Fe\ione\ and  Fe\ii\ lines in the non-LTE analysis. 

\begin{figure}
	\includegraphics[width=80mm]{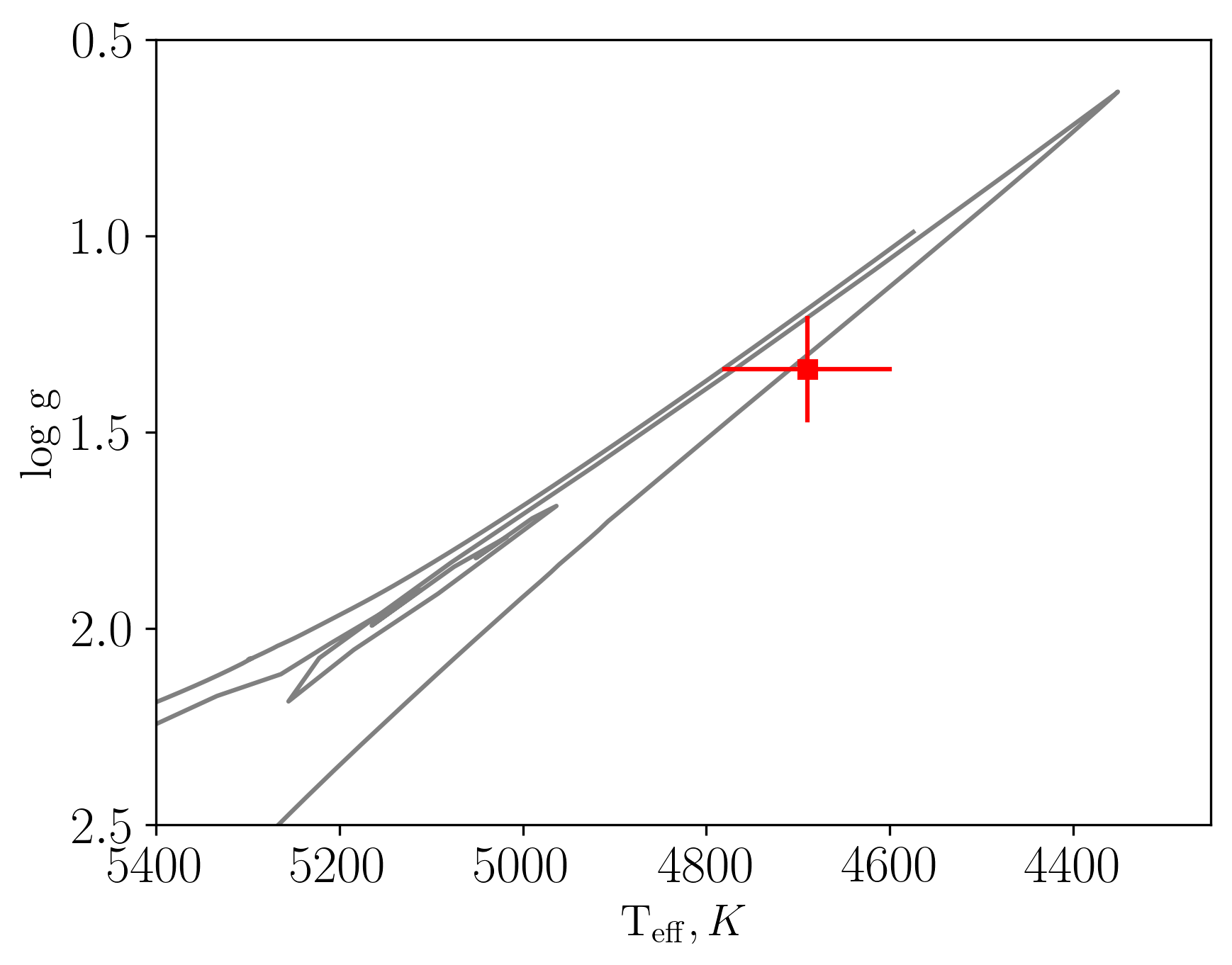}
	\caption{The position of the star of interest (red square) on the \teff\--log~g diagram and the evolutionary track with the corresponding parameters from \citet{2016ApJS..222....8D} grid.}
	\label{track} 
\end{figure}

Using the derived stellar parameters we compared the position of the star of interest at the \teff\--log~g diagram with the corresponding evolution track from \citet{2016ApJS..222....8D} grid (Fig.~\ref{track}). The star sits well on the red giant branch and its parameters correspond to age of 12.2~Gyr. Considering the uncertainties in \teff\ and log~g, we cannot exclude that the star may belong to a more advanced evolutionary stage.

\section{Abundance Analysis}
\label{abund}

\subsection{Codes and model atmospheres}

We use classical 1D model atmospheres from the \textsc {marcs} model grid \citep{marcs}, interpolated for the given \teff, log~g, and [Fe/H] of the star.

We solve the coupled radiative transfer and statistical equilibrium equations with the \textsc {detail} code \citep{detail}, using the updated opacity package as presented by  \citet{mash_fe}. For synthetic spectra calculations, we use the \textsc {synthV\_NLTE} code \citep{Tsymbal2018} attached to the \textsc {idl binmag} code  \citep{2018ascl.soft05015K}. This technique allows us to obtain the best fit to the observed line profiles with the non-LTE effects taken into account via  pre-calculated departure coefficients (the ratio between non-LTE and LTE atomic level populations) for a given model atmosphere.
When fitting the line profiles, the abundance of the element of interest is varied together with the macroturbulent velocity (v$_{\rm mac}$) and the radial velocity (v$_{\rm r}$). For the star of interest, the typical uncertainties in these parameters caused by the fitting procedure are 0.03 dex, 0.5 \kms, and 0.1 \kms, respectively. These correspond to uncertainties on individual lines, while the real uncertainties caused by a scatter between different lines are larger. Using our linelist, we calculate uncertainties of 2.5 \kms\ and 0.8 \kms\ in v$_{\rm mac}$ and v$_{\rm r}$, respectively.

The line list for spectral synthesis is extracted from a recent version of the Vienna Atomic Line Database \citep[VALD,][]{2019ARep...63.1010P,2015PhyS...90e4005R} that provides isotopic and hyperfine
structure components of the spectral lines for a number of chemical elements. For lithium, the data on the fine and hyperfine structures and isotope shifts originate from \citet{1995PhRvA..52.4462R}. VALD provides a linelist computed for solar isotopic ratios. To determine the $^6$Li/$^7$Li isotopic ratio, we rescale the original data adopting different isotopic ratios.

\subsection{Non-LTE effects}

We take into account the departure from LTE for a number of chemical elements (Li, Na, Mg, Ca, Ti, Cr, Mn, Fe, Zn, Sr, Ba). We refer the reader to the papers listed in Table~\ref{tab:atoms} for the description of the model atoms and the mechanism of the non-LTE effects. 
For most chemical elements, we perform non-LTE calculations with the specific model atmosphere, while for Na\ione, Cr\ione, Mn\ione, and Zn\ione, we   interpolate the non-LTE corrections ($\Delta_{\rm NLTE}$ = $\eps_{\rm NLTE}$ -- $\eps_{\rm LTE}$) in the pre-calculated grids available in the literature. 
Our manganese abundance relies on the  Mn\ione\ 4783 \AA\ line only. For this line, the updated grid of non-LTE corrections on the MPIA webpage\footnote{https://nlte.mpia.de/gui-siuAC\_secE.php} does not cover our stellar parameters. From the previous version of the grid, we derive \dnlte = 0.52 and 0.58~dex for the Mn\ione\ 4783 and 4823 \AA\ lines, respectively. The updated grid provides \dnlte = 0.34~dex for the Mn\ione\  4823 \AA\ line. We assume the non-LTE corrections are similar for these two lines and adopt \dnlte = 0.34~dex for the Mn\ione\ 4783 \AA\ line.

For the remaining chemical elements (Si, Sc, Ni Y), the non-LTE effects are either small or unavailable in the literature for the stellar parameters investigated in this study.
For Si\ione, non-LTE effects are minor and can be neglected even in metal-poor stars \citep{2016AstL...42..366M}. For Sc\ii, \citet{2022AstL...48..455M} investigated the departures from LTE  and found positive non-LTE abundance corrections in metal-poor dwarfs. However, we cannot apply their results obtained for dwarfs to our giant star. 
The departures from LTE for Ni\ione\ were investigated in solar atmosphere by \citet{1993AA...269..509B,2013ApJ...769..103V,2021MNRAS.508.2236B,2022A&A...661A.140M} and in FGK stars by \citet{eitnerni}. For Ni\ione\ lines in the visible range, \citet{eitnerni} predict positive non-LTE abundance corrections, which increase towards higher \teff\ and lower log~g. For example, they found $\Delta{\rm NLTE}$ = 0.2~dex in model atmosphere with \teff/log~g/[Fe/H] = 5000/3/--2.5. 
For  Y\ii\ 4883 and 5205 \AA\ lines, we applied non-LTE abundance corrections of $\Delta_{\rm NLTE}$ = 0.14~dex computed by \citet{2023arXiv230901402A} for model atmosphere with \teff/log~g/[Fe/H] = 5000/2.0/--2.5.

\begin{table}
	\caption{References for the  non-LTE methods used in this study}
	\label{tab:atoms}
	\begin{tabular}{ll} 
		\hline
	Species & Reference   \\
	\hline
Li\ione        & this study  \\
Na\ione        & \citet{2022AA...665A..33L} \\ 
Mg\ione        & \citet{2013AA...550A..28M}  \\ 
Ca\ione\       & \citet{2017AA...605A..53M}  \\
Ti\ione--\ii\  & \citet{2020AstL...46..120S}  \\
Cr\ione\       & \citet{2010AA...522A...9B}  \\
Mn\ione\       & \citet{2019AA...631A..80B}  \\
Fe\ione\--\ii\ & \citet{mash_fe}  \\
Zn\ione\       & \citet{2022MNRAS.515.1510S}  \\
Sr\ii\         & \citet{atoms10010033}  \\
Y\ione--\ii\   & \citet{2023arXiv230901402A}  \\
Ba\ii\         & \citet{2019AstL...45..341M} \\
		\hline
   	\end{tabular}
\end{table}

\begin{table}
	\caption{Non-LTE and LTE abundance ratios}
\setlength{\tabcolsep}{1.10mm}
 \label{abund_table}
	\begin{tabular}{lllrlrl} 
		\hline
Species & $\eps_{\rm \odot}$ & [X/H]  & [X/FeII] & [X/H] & [X/FeII] & N \\ 
 &  & LTE  & LTE & NLTE & NLTE &  \\ 
	\hline
   Li\ione  & 1.05 &   2.23              &   4.66 &   2.39               &   4.82 &  1 \\
     CH     & 8.39 & --2.83 $\pm$ 0.12 & --0.40 &                      &        &  1 \\
   O\ione   & 8.73 & $<$ --1.93          & $<$ 0.50 &  $<$ --1.93            & $<$ 0.50 &  1 \\     
   Na\ione  & 6.29 & --2.01   $\pm$ 0.01 &   0.42 &  --2.36 $\pm$   0.03 &   0.07 &  2 \\
   Mg\ione  & 7.54 & --2.15   $\pm$ 0.24 &   0.28 &  --2.15 $\pm$   0.19 &   0.29 &  2 \\
   Si\ione  & 7.53 & --2.13              &   0.30 &                      &        &  1 \\
   Ca\ione  & 6.31 & --2.15   $\pm$ 0.14 &   0.28 &  --2.09 $\pm$   0.13 &   0.35 &  7 \\
   Sc\ii    & 3.07 & --2.42   $\pm$ 0.02 &   0.02 &                      &        &  3 \\
   Ti\ii    & 4.93 & --2.11   $\pm$ 0.11 &   0.33 &  --2.08 $\pm$   0.11 &   0.35 &  6 \\
   Cr\ione  & 5.65 & --2.67              & --0.24 &  --2.25 $\pm$   0.07 &   0.18 &  1 \\
   Mn\ione  & 5.50 & --3.15              & --0.72 &  --2.81 $\pm$   0.07 & --0.38 &  1 \\
   Fe\ione  & 7.46 & --2.50   $\pm$ 0.16 & --0.06 &  --2.45 $\pm$   0.16 & --0.02 & 57 \\
   Fe\ii    & 7.46 & --2.43   $\pm$ 0.07 &   0.00 &  --2.43 $\pm$   0.07 &   0.00 &  6 \\
   Ni\ione  & 6.22 & --2.52   $\pm$ 0.01 & --0.09 &                      &        &  2 \\
   Zn\ione  & 4.65 & --2.36              &   0.07 &  --2.21 $\pm$   0.07 &   0.22 &  1 \\
   Sr\ii    & 2.90 & --2.54   $\pm$ 0.00 & --0.11 &  --2.52 $\pm$   0.01 & --0.09 &  2 \\
   Y\ii     & 2.20 & --2.73   $\pm$ 0.02 & --0.29 &  --2.59 $\pm$   0.02 & --0.15 &  2 \\
   Ba\ii    & 2.18 & --3.06   $\pm$ 0.15 & --0.62 &  --3.05 $\pm$   0.18 & --0.62 &  4 \\
 \hline
   	\end{tabular}
The abundance uncertainty is calculated as the dispersion of the single line measurements around the mean
$\sigma = \sqrt{ \Sigma (\eps - \eps_i )^2 /(N - 1)}$, where N is the total number of lines.    
\end{table}

\begin{table}
\caption{NLTE and LTE abundances from individual lines and their atomic data.}
\label{tab:individual}
\begin{tabular}{llcrrrr} 
\hline
Sp.  &  $\lambda$,  & \eexc, & loggf & EW,  & $\eps$ & $\eps$ \\
     &  \AA\        &  eV    &        &  m\AA & LTE  & NLTE   \\
\hline
 Li\ione\ & 6103.65 & 1.85 &   0.58 &  71.5  & 3.27 &   3.44 \\
 Li\ione\ & 6707.91 & 0.00 &   0.17 & 426.0  & --   &   3.44 \\ 
 CH       & 4313.00 & --   &    --  &    --  & 5.56 &   --   \\ 
 O\ione\  & 6300.30 & 0.00 &  --9.78 &   6.6  & $<$ 6.80 &  $<$ 6.80 \\
 Na\ione\ & 5889.95 & 0.00 &    0.11 & 228.4  & 4.28 &   3.95 \\
 Na\ione\ & 5895.92 & 0.00 &  --0.19 & 201.7  & 4.29 &   3.91 \\
\hline 
\end{tabular}\\
This table is available in its entirety on the last page. A portion is shown here for guidance regarding its form and content.
\end{table}  

\subsection{Lithium abundance determination}\label{li_atom}

\subsubsection{Li\ione\ model atom}\label{atom}

\begin{figure}
	\includegraphics[width=80mm]{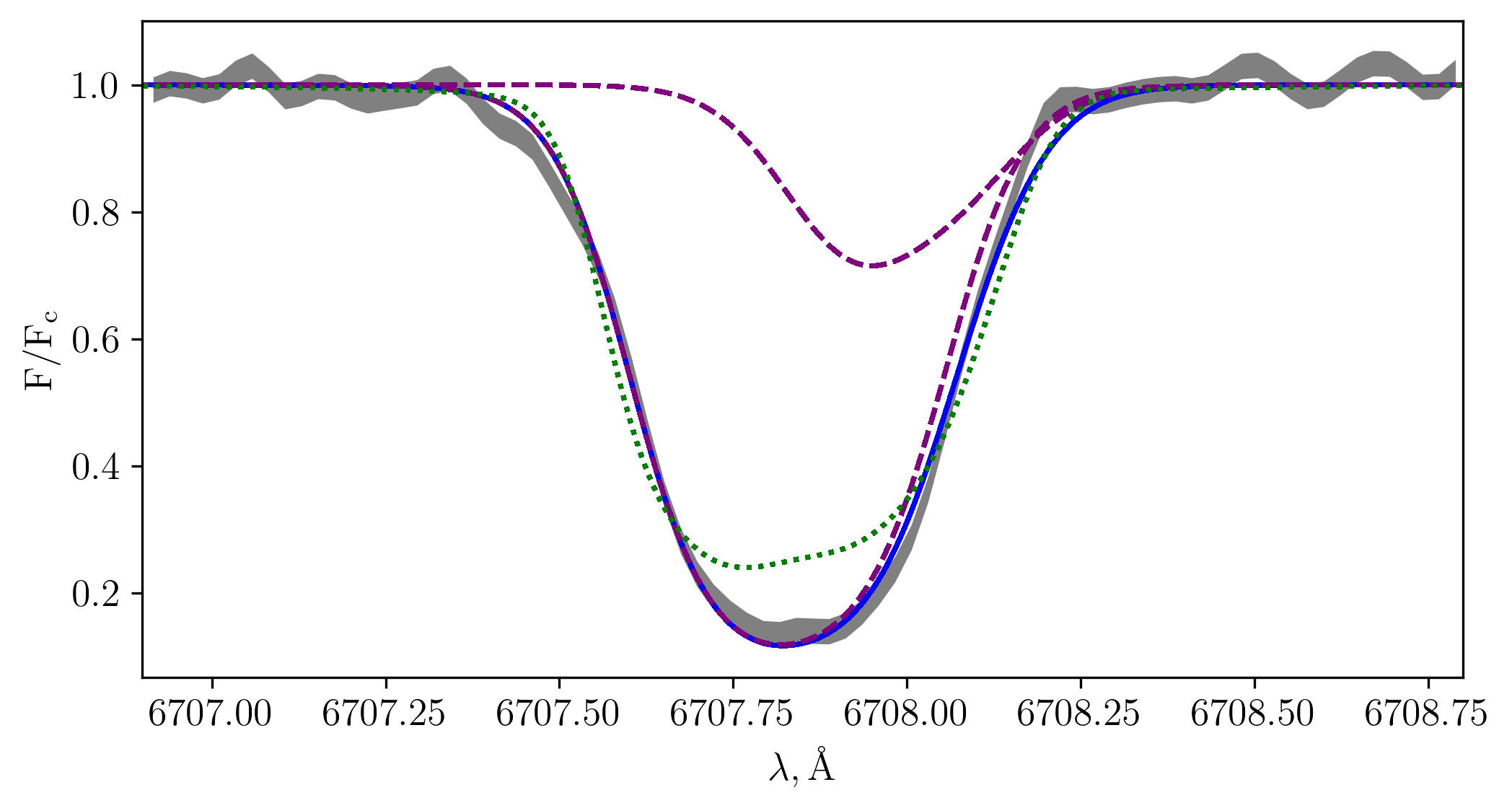}
    \includegraphics[width=80mm]{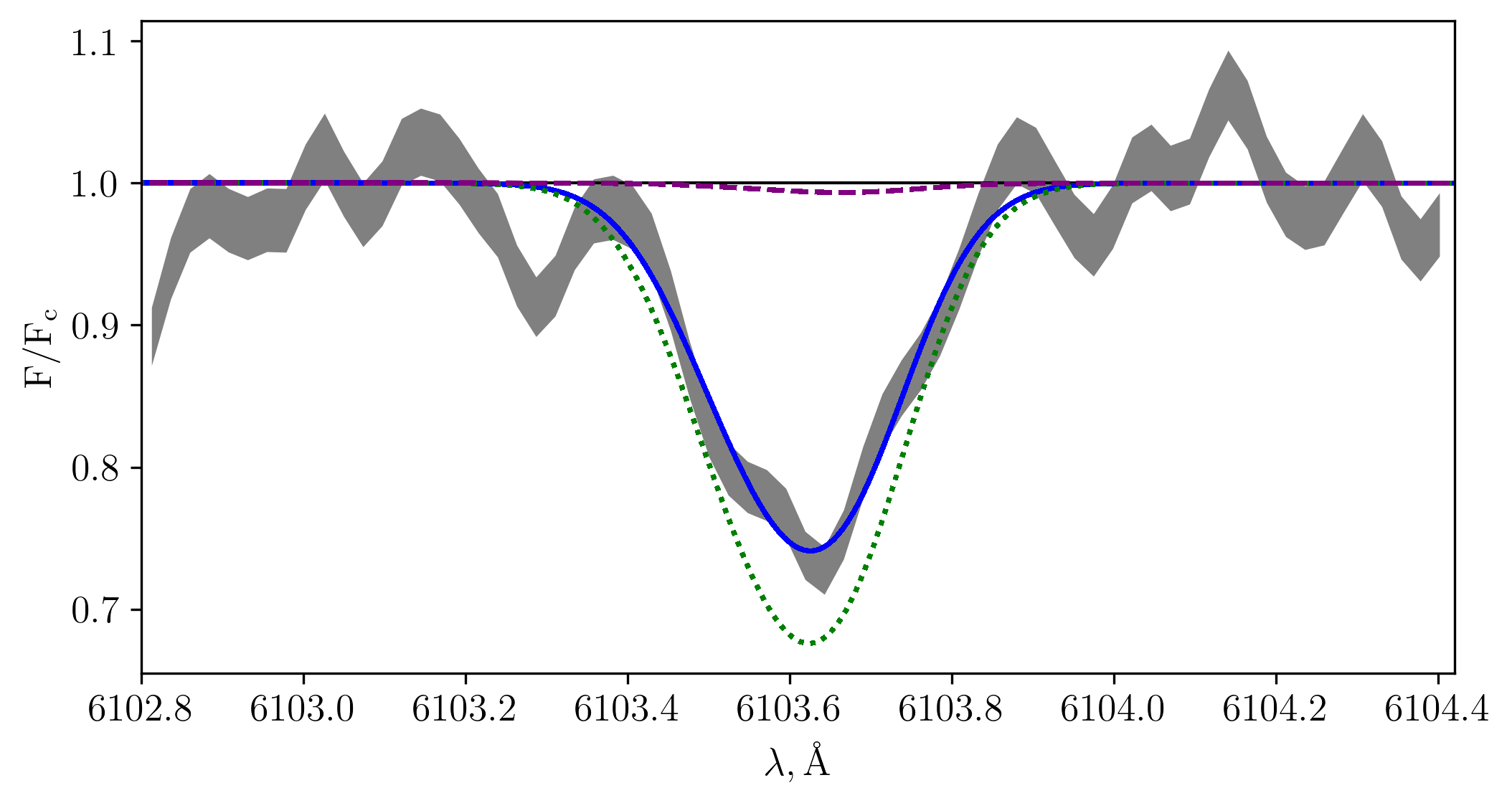}
	\caption{Top panel: Best fit of the Li\ione\ 6707 \AA\  non-LTE 1D line profile (solid curve), together with the best fit LTE profile (dotted curve). The non-LTE and LTE fits correspond to $\eps$ = 3.44 and 3.84, respectively. Dashed lines show the contribution from the $^7$Li and $^6$Li to the non-LTE profile assuming  $^6$Li/$^7$Li = 1.67\%. Bottom panel: Best fit of the Li\ione\ 6103 \AA\ non-LTE (solid curve) and LTE line profiles (dotted curve), computed for the same abundance $\eps$ = 3.44. The observed spectrum and associated uncertainties of the star of interest are represented by the shaded area. }
	\label{6707_nlte_lte} 
\end{figure}

In the studied star,  the Li\ione\ 6707 \AA\ resonance line is strong with EW = 426 m\AA\ and its profile cannot be fitted in LTE with any abundance and macroturbulent velocity. Non-LTE leads to a strengthened core of the Li\ione\ 6707 \AA\ line and thus allows us to fit the observed profile (see Fig.~\ref{6707_nlte_lte}).
To account for the non-LTE effects, we construct a  Li\ione\ model atom. It includes 21 levels of Li\ione\ and the ground state of Li\ii. The list of energy levels and transitions is taken from R.~Kurucz's webpage\footnote{http://kurucz.harvard.edu/atoms.html}.
We included  all levels in the model atom, up to the ionization threshold of 5.39~eV. Levels with an excitation energy larger than 5~eV are combined into the six superlevels according to their parity. In the statistical equilibrium calculations, we neglect the fine structure and the $^6$Li isotope. 
We adopt photoionization cross-sections from the R-matrix calculations of \citet{1988JPhB...21.3669P}, available in the TOPbase\footnote{http://cdsweb.u-strasbg.fr/topbase/topbase.html}. Inelastic collisions with hydrogen atoms are taken from \citet{2003A&A...409L...1B}.
Electron impact excitation rates were taken from quantum-mechanic calculations of \citet{2012JPhCS.388d2018O}, where available.
For the remaining radiatively allowed and forbidden transitions, electronic collision rates are calculated with the approximate formulae from \citet{Reg1962} and \citet{1948MNRAS.108..292W}, respectively. Electron impact ionisation rates are calculated with the \citet{1962amp..conf..375S} formula.  

As a sanity check of our model atom, we compare our non-LTE results  with those calculated by  \citet{2007A&A...465..587S} and \citet{2009AA...503..541L} with their original model atoms.
We compute the non-LTE abundance corrections for the Li\ione\ 6707 \AA\ line in a model atmosphere with \teff/log~g/[Fe/H]/\vt = 5806/3.69/$-2.42$/1.5 and $\eps_{\rm Li}$ = 2.2. For this model, \citet{2007A&A...465..587S} and \citet{2009AA...503..541L} provide \dnlte = 0.05 and $-0.05$ dex, respectively. Our non-LTE calculations agree with  \citet{2007A&A...465..587S} and we find the same \dnlte = 0.05. 
Here, the  \dnlte\ of  \citet{2009AA...503..541L} was derived by interpolation on a grid of non-LTE corrections. If we interpolate  \dnlte\ for the resonance line for solar model atmosphere and a metal-poor giant with 4500/1.5/$-2.0$/2.0 and [Li/H] = 0, we find \dnlte = 0.05 and 0.15 dex, respectively. Our non-LTE corrections are 0.1 dex smaller for both model atmospheres: \dnlte = $-0.05$ and 0.05 dex. 
This discrepancy can be explained by the recent findings of \citet{2021MNRAS.500.2159W} that non-LTE corrections are ``up to 0.15~dex more negative than in previous work'' due to the incorrect accounting of the Li\ione\ ultraviolet lines in \citet{2009AA...503..541L}. 

\subsubsection{3D effects}\label{3deffect}

\begin{figure}
	\includegraphics[width=80mm]{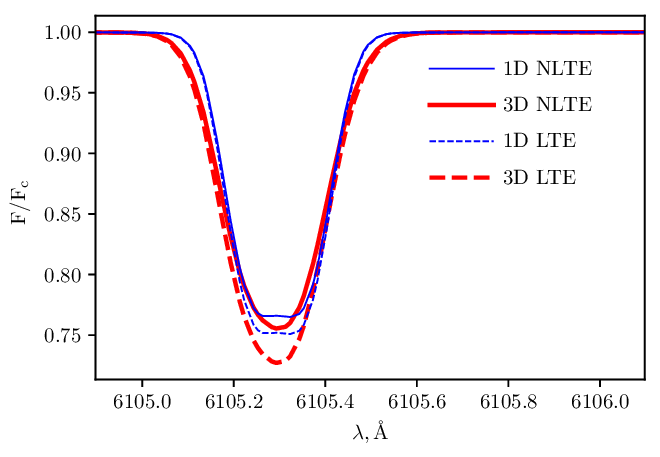}
 	\includegraphics[width=80mm]{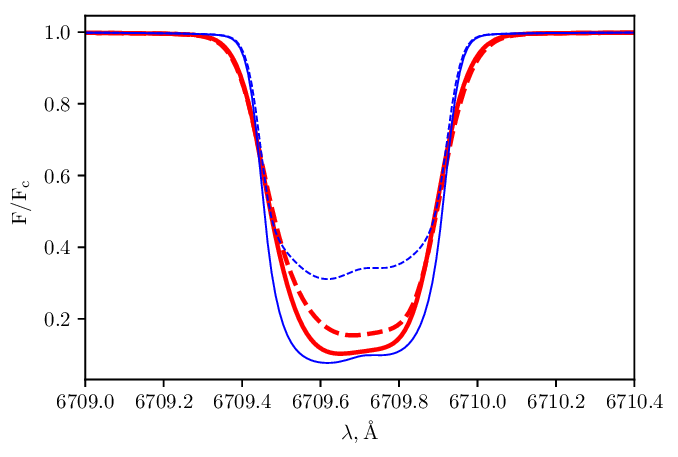}
	\caption{Li\ione\ 6103 and 6707 \AA\ line profiles for different line formation scenarios, extracted from the BREIDABLIK package for a model atmosphere with 4500/1.5/$-2.0$ and $\eps_{Li}$ = 3.0. See the legend for the designations.}
	\label{3dplots} 
\end{figure}

To estimate the impact of the 3D effects on the Li\ione\ 6103 and 6707~\AA\ lines, we adopt synthetic spectra grid computed by \citet{2021MNRAS.500.2159W} and the BREIDABLIK package\footnote{https://github.com/ellawang44/Breidablik}. 
Fig.~\ref{3dplots} shows the Li\ione\ 6103 and 6707 \AA\ line profiles extracted for different line formation scenarios (1D LTE, 3D LTE, 1D NLTE, 3D NLTE) in a model atmosphere with 4500/1.5/$-2.0$ and $\eps_{Li}$ = 3.0. 
For the subordinate Li\ione\ 6103 \AA\ line, the 3D NLTE profile is slightly stronger than the 1D NLTE one, which translates to a $-0.04$~dex abundance difference. For the resonance Li\ione\ 6707~\AA\ line, the 3D NLTE calculation results in slightly stronger wings, but a weaker line core compared to the 1D NLTE case, which results in a $0.14$~dex abundance difference between 3D NLTE and 1D NLTE. 

We extract 1D~non-LTE and 3D~non-LTE synthetic spectra from BREIDABLIK for the node models around the stellar parameters of the star of interest. Integrating the synthetic profiles and using the growth curve we compute the 3D abundance corrections for each node model that were used for the interpolation of the 3D corrections, assuming the atmospheric parameters of our star (Table~\ref{tab:3d}). We interpolate 3D corrections in \teff, [Fe/H], and $\eps_{Li}$. The surface gravity is fixed with log~g = 1.5 since the BREIDABLIK grid does not contain synthetic spectra for models with log~g $<$ 1.5. To estimate  the impact of log~g on 3D corrections, we provide test calculations for model with 4500/2.0/$-2.0$ and $\eps_{Li}$ = 3.5 (Table~\ref{tab:3d}). 
Finally, based on the Li\ione\ 6707 \AA\ and 6103 \AA\ lines of the spectrum, we calculate $\Delta_{\rm 3D} = -0.17$ and 0.02 dex, respectively. 

It is worth noting that the calculations of \citet{2021MNRAS.500.2159W} account for only the primary lithium isotope, $^7$Li.
Given that the contribution of the $^6$Li isotope is two orders of magnitude smaller than that of $^7$Li, and considering that we use the synthetic spectra of \citet{2021MNRAS.500.2159W} for a self-consistent determination of 3D abundance corrections by comparing their 3D NLTE and 1D NLTE profiles, we assume that the impact of $^6$Li on the 3D corrections is small compared to the 3D corrections themselves. Further 3D NLTE calculations with different lithium isotopic ratios would be helpful in this regard.

Here, we speculate on the impact of $^6$Li on the 3D abundance corrections for the Li\ione\ 6707 \AA\ line. As shown in Fig.~\ref{3dplots} (bottom panel),
the wings of the 3D profile are stronger, while the core appears weakened compared to the 1D profile. This implies that, on average, the 3D model atmosphere is cooler in deep layers and hotter in high layers compared to the 1D model. Strengthened and weakened line profiles result in negative and positive 3D abundance corrections, respectively. Depending on the Li\ione\ 6707 \AA\ line strength, the corresponding 3D corrections may have different sign and absolute value depending on a contribution from the wings and the core, i.e. line formation depths.

Including $^6$Li results in a broader profile and a larger contribution from deep atmospheric layers to the total profile. In other words, when both isotopes are considered, the line will, on average, form at deeper atmospheric layers. Thus, including $^6$Li will lead to a smaller 3D abundance correction due to the larger contribution from the wings. Smaller 3D correction for the Li\ione\ resonance line results in a smaller $^6$Li/$^7$Li ratio. Thus, our 3D $^6$Li/$^7$Li ratio can be considered as an upper limit.

\begin{table}
\caption{3D abundance corrections for the node model atmospheres}
\label{tab:3d}
\begin{tabular}{llllll} 
\hline
\teff & log g & [Fe/H] & $\eps_{Li}$ & $\Delta_{\rm 3D, 6707}$ & $\Delta_{\rm 3D, 6103}$ \\
\hline
4500 &  1.5 & --3.00 & 3.00 & 0.09 & --0.02 \\
4500 &  1.5 & --3.00 & 3.50 & 0.15 & --0.01 \\
4500 &  1.5 & --2.00 & 3.00 & 0.14 & --0.04 \\
4500 &  1.5 & --2.00 & 3.50 & 0.20 & --0.03 \\
4500 &  2.0 & --2.00 & 3.50 & 0.31 & --0.05 \\
4750 &  1.5 & --3.00 & 3.00 & 0.09 & --0.02 \\
4750 &  1.5 & --3.00 & 3.50 & 0.15 & --0.01 \\
4750 &  1.5 & --2.00 & 3.00 & 0.14 & --0.04 \\
4750 &  1.5 & --2.00 & 3.50 & 0.20 & --0.03 \\
\hline
 4690 & 1.5	& --2.42 & 3.44 & 0.17 & --0.02 \\
\hline
\end{tabular}
\end{table}

\subsubsection{Testing the method with reference star HD~140283}

Our measurement of a lithium isotopic ratio assumes that the two Li\ione\ lines yield consistent lithium abundances when we infer the correct isotopic ratio (see Sect.~\ref{67ratio}).
In order to test this assumption,  we first analyse the two Li\ione\ lines of unevolved low-metallicity stars with normal lithium abundances.
For these stars on the Spite plateau, 3D non-LTE analyses of the Li\ione\ 6707 \AA\ line profiles showed that there is no detectable $^6$Li in their atmospheres \citep{2013A&A...554A..96L,2022MNRAS.509.1521W,2019A&A...628A.111G}. 
The goal of our test is to verify if we get consistent abundances from Li\ione\ 6103 and 6707 \AA\ lines for these stars with $^6$Li/$^7$Li$=0$.
We also note that their Li\ione\ 6707 \AA\ lines are weak and the lithium isotope ratio does not affect their strength.

We first test our lithium abundance determination method with the well-studied metal-poor star HD~140283. 
We adopt \teff/log~g/[Fe/H]/\vt\ = 5780/3.70/2.38/--2.43/1.3, with its \teff\ and log~g taken from \citet{lick} and in agreement with \teff\ = 5787 $\pm$ 48~K, as measured by \citet{2018MNRAS.475L..81K}, and log~g = 3.66 $\pm$ 0.03, calculated using the Gaia parallax \citep{2021A&A...649A...1G}. The metallicity and the microturbulent velocity are taken from the non-LTE analysis of iron lines performed by \citet{2019AA...631A..43M}.

We use a high-resolution and high S/N spectrum of HD~140283 taken on 27 January 2017 with the same instrument (Subaru/HDS) and reduced using the same procedure as for the star of interest (see Sect.~\ref{parameters}). We calculate signal-to-noise ratios of S/N = 1080 and 1090 around the Li\ione\ 6103 \AA\ and 6707 \AA\ lines, respectively. This high quality spectrum ensures reliable line profile fitting and gives EW = 1.6 m\AA\ and 47.8 m\AA\ for the subordinate and the resonance line, and the corresponding uncertainties in abundances of 0.09~dex and 0.01~dex. In 1D non-LTE, Li\ione\ 6103 \AA\ and 6707 \AA\ give $\eps_{\rm NLTE}$ = 2.22 and 2.27, respectively. For both Li\ione\ lines in HD~140283, 3D effects lead to weakened lines and positive abundance corrections: $\Delta_{\rm 3D}$ = 0.07 and 0.09 for Li\ione\ 6103 \AA\ and 6707 \AA, respectively. Thus, for HD~140283, the two lines give consistent results within the uncertainties for the non-LTE abundances, either in 1D or after applying the 3D abundance corrections. To make sure that HD~140283 is not an isolated case, we apply our non-LTE method for lithium abundance determination from the Li\ione\ 6103 \AA\ and 6707 \AA\ lines on a sample of stars.

\subsubsection{Testing the method with a sample of MP stars}
\begin{figure}
	\includegraphics[width=80mm]{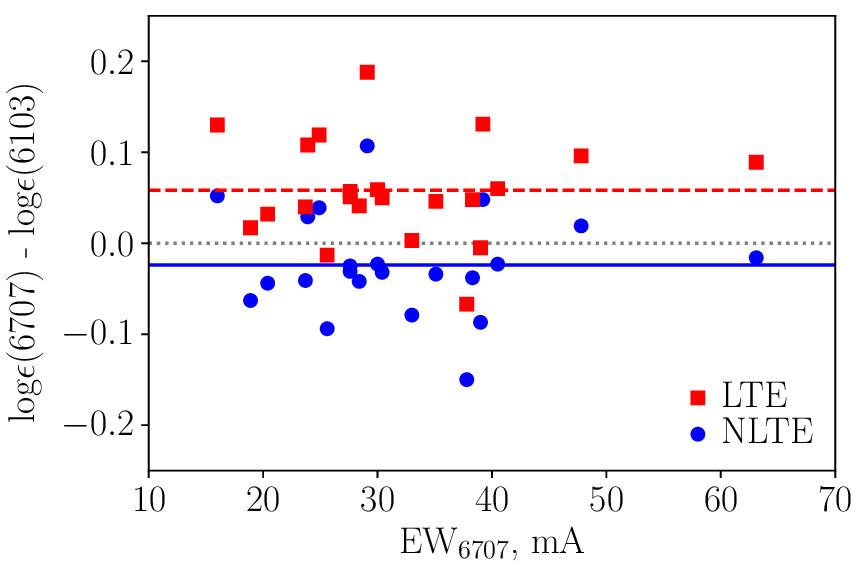}
	\includegraphics[width=80mm]{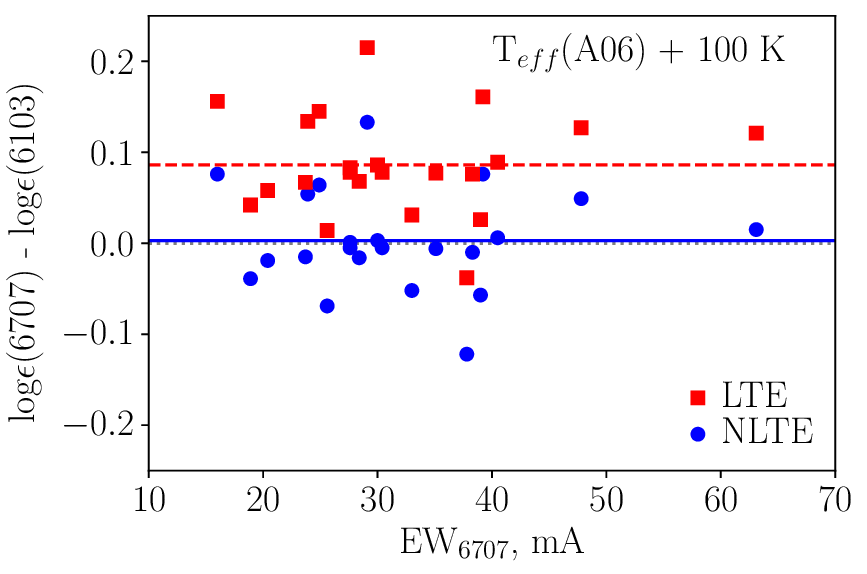}
	\caption{Non-LTE (blue circles) and LTE (red squares) abundance differences between the Li\ione\ 6707 \AA\ and 6103 \AA\ lines in a sample of MP stars with normal lithium abundances.  For comparison, the bottom panel shows the same as in the top panel, but with \teff\ increased by 100~K. }
	\label{a06} 
\end{figure}

We further test our assumption using 22 stars on the Spite plateau from \citet[][hereafter, A06]{2006ApJ...644..229A}, which provided stellar parameters and equivalent widths for the 6103 and 6707 \AA\ lines.
We rederive lithium abundances from each of the two lines adopting the stellar parameters from A06 and 
Fig.~\ref{a06} shows the non-LTE and LTE abundance differences between the two Li\ione\ lines. In LTE, the Li\ione\ 6707 \AA\ line gives, on average, 0.06~dex higher abundance compared to the 6103 \AA\ line, while they are $-0.02$~dex lower in non-LTE. Thus, non-LTE reduces the abundance difference between the two lines.

A06 determined \teff\ from fitting the H$_{\alpha}$ wings in LTE but \citet{2008A&A...478..529M} found that fitting the wings of the Balmer lines in non-LTE leads to $\sim$60~K higher \teff\ compared to the LTE case. 
\citet{2018A&A...615A.139A} found that, in metal-poor turn-off stars, \teff\ determined from the wings of H$_{\alpha}$ in 3D non-LTE is 150 K higher compared to 1D LTE. A systematic uncertainty in \teff\ results in a systematic discrepancy in abundances between the two lines, since the resonance line is more affected by changes in \teff\ compared to the subordinate line (see Table~\ref{tab:lierr}).
As a test, we increase the effective temperatures of A06 by 100~K and calculate the non-LTE and LTE lithium abundances. 
Fig.~\ref{a06} (bottom panel) presents the abundance differences between the two Li\ione\ lines derived with the increased \teff. 
In non-LTE, the two lines give consistent abundances, and, on average, the abundance difference $\Delta(6707-6103)_{\rm NLTE}$ = 0.00 $\pm$ 0.06, while the LTE assumption results in a larger abundance from the resonance line and $\Delta(6707-6103)_{\rm LTE}$ = 0.09 $\pm$ 0.06. 

In conclusion, our test calculations show that, in different VMP stars, non-LTE leads to consistent abundances from the Li\ione\ 6707 \AA\  and 6103 \AA\ lines.

\section{Abundances}\label{results}

In total, we derive abundances for 16 chemical elements in the star of interest. Our average LTE and non-LTE abundance are presented in Table~\ref{abund_table}, abundances from individual lines together with their equivalent width and atomic data are given in Table~\ref{tab:individual}.
Carbon abundance determination is based on the CH 4300 \AA\ G band and its error is estimated by applying a continuum placement shift, which results in [C/Fe] = --0.40 $\pm$ 0.12. Accounting the luminosity of the star of interest log(L/L$_{\odot}$) = 2.6, the derived [C/Fe] is typical for VMP stars with normal carbon abundance according to \citet{2007ApJ...655..492A} classification.

Our observed spectrum covers the [O\ione ] 6300 \AA\ forbidden line wavelength range, however, this line is not detected. Using this spectral region and applying the macroturbulent velocity v$_{\rm mac}$ = 5.1 \kms, we estimated an upper limit [O/Fe] $<$ 0.50. The forbidden line is immune to the departures from LTE, thus, we do not perform non-LTE calculations for O\ione. A typical [O/Fe] ratio in very metal-poor stars is [O/Fe] = 0.6 \citep[see, for example,][]{2004A&A...416.1117C}. Although the derived upper limit is lower compared to the typical ratio, a spectrum with higher signal to noise ratio is required to prove that the star of interest indeed has low oxygen abundance.

Different $\alpha$-elements (Mg, Si, Ca) and Ti show similar  [$\alpha$/Fe] ratios ($\sim$0.3).  Scandium, chromium, nickel and zinc abundances follow iron within the uncertainties. For manganese, we find [Mn/Fe] = $-0.4$, which is in line with the non-LTE trend found by \citet{2022arXiv220610258E} for halo stars. Neutron-capture elements are represented by Sr, Y, and Ba. In non-LTE we find [Sr/Ba] = 0.5 and [Ba/H] = $-3.0$, which is in line with expectations for a typical metal-poor star in the MW halo \citep[see, for example,][]{2017A&A...608A..89M}.

\begin{figure}
	\includegraphics[width=80mm]{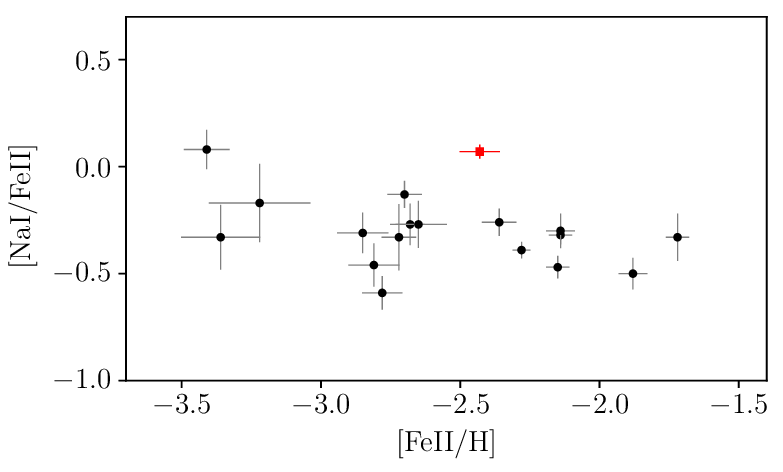}
	\caption{Non-LTE [Na/Fe] ratios in the star of interest (red square) and comparison sample giants (black circles) from \citep{2017A&A...608A..89M}. }
	\label{nafe_feh} 
\end{figure}

The above element abundances are similar for those measured in typical metal-poor halo stars with similar [Fe/H]. The exceptions are lithium and sodium. In non-LTE, we calculate [Na/Fe] = 0.07 $\pm$ 0.03, which is higher compared to [Na/Fe] = $-0.4$ found in non-LTE by \citet{2017A&A...608A..89M} for stars with similar [Fe/H] (Fig.~\ref{nafe_feh}). 

The star is strongly enhanced in lithium such that the subordinate Li\ione\ 6103 \AA\ line is clearly detected (Fig.~\ref{6707_nlte_lte}). It has EW = 72 m\AA\ and gives $\eps$ = 3.44 in non-LTE regardless of the adopted $^6$Li/$^7$Li ratio. The resonance Li\ione\ 6707 \AA\ line is strong with EW = 426 m\AA\ and, in contrast to the subordinate line, it is sensitive to  the isotopic ratio. 

\section{$^6$Li/$^7$Li isotopic ratio}\label{67ratio}

\begin{figure}
	\includegraphics[width=80mm]{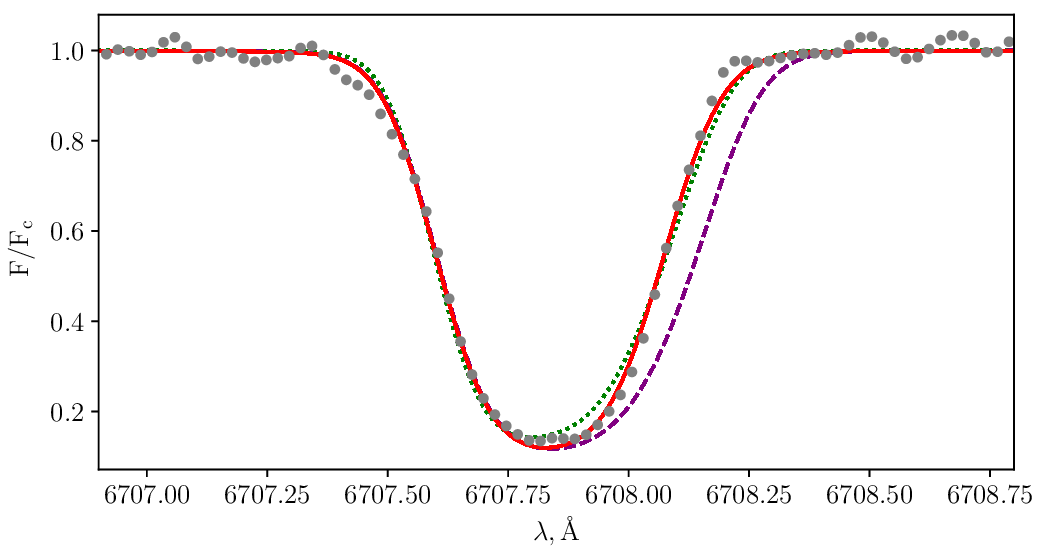}
	\caption{
Li\ione\ 6707 \AA\ non-LTE 1D line profiles calculated with $^6$Li/$^7$Li = 0.5~\% and $\eps$ = 3.54 (red solid curve); $^6$Li/$^7$Li = 8 \% and log$\epsilon$ = 3.54 (purple dashed curve); $^6$Li/$^7$Li = 8 \% and $\eps$ = 3.19 (green dotted curve). The observed spectrum of the investigated star is shown with dots.
}
	\label{6707profile} 
\end{figure}

\begin{figure}
 	\includegraphics[width=80mm]{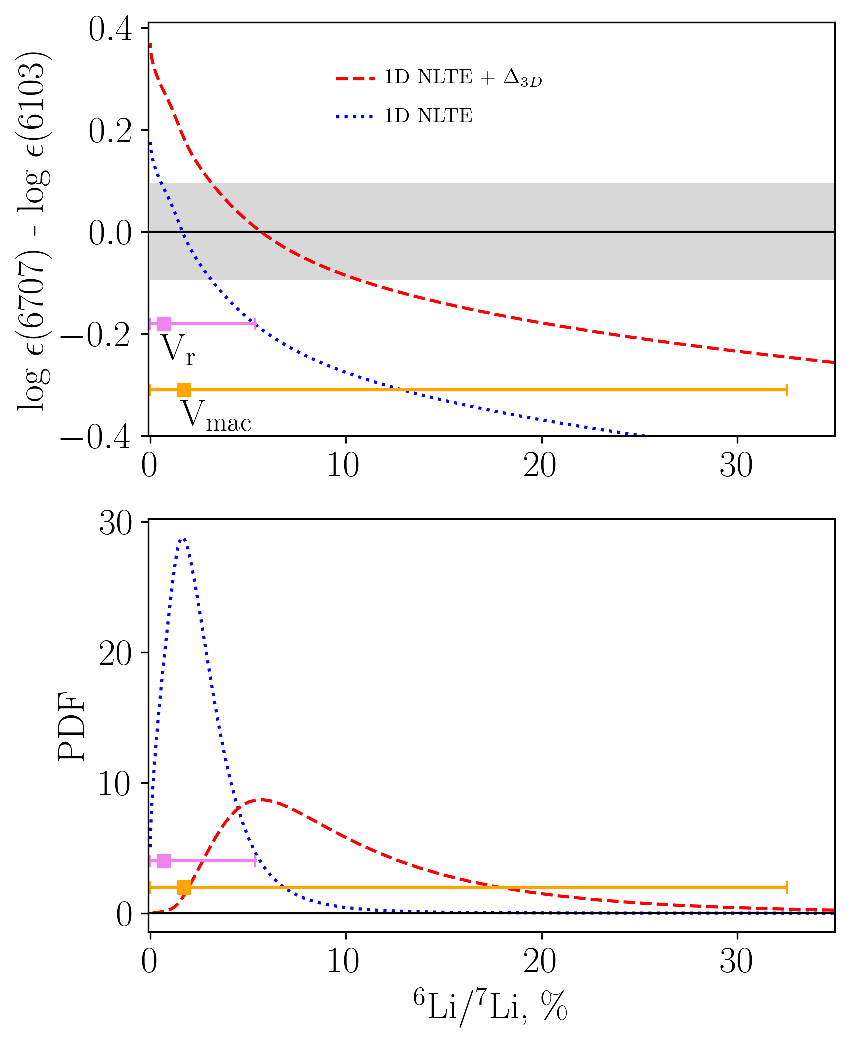}
\caption{Top panel: Non-LTE abundance difference between the Li\ione\  6707 and 6103 \AA\ lines in 1D (dotted curve) and after applying 3D corrections (dashed curve) as a function of $^6$Li/$^7$Li isotopic ratio. The uncertainty in abundance difference of 0.1~dex is shown with the shaded area. Bottom panel: Probability distribution functions in 1D (dotted curve) and after applying 3D corrections (dashed curve) as a function of $^6$Li/$^7$Li isotopic ratio. 
Horizontal lines indicate limitations on $^6$Li/$^7$Li from analysis of v$_{\rm r}$ and v$_{\rm mac}$ of the best fit profiles.}
	\label{abiso} 
\end{figure}

We determine the lithium isotope ratio from the  Li\ione\ 6707 \AA\ line. 
Our fitting procedure for this line is similar to what we adopted for the other lines in the spectrum, which yields the $\eps_{Li}$ that minimizes the difference between the observed and synthetic spectra.
We fit the resonance line adopting different $^6$Li/$^7$Li isotope ratios from 0 to 50 \%\ (Table~\ref{tab:67ratio}).
For each of the ratios, we vary the  abundance, $\eps_{Li}$, and adjust line position and width. 
Since the strength of this line depends on both the lithium isotope ratio and the lithium abundance, we obtain different best-fit abundances when different isotope ratios are assumed (see Fig.~\ref{6707profile}). Table~\ref{tab:67ratio} lists the parameters ($\eps_{Li}$, v$_{\rm mac}$, and v$_{\rm r}$) of the bestfit synthetic spectra of the Li\ione\ 6707 \AA\ computed for different $^6$Li/$^7$Li isotope ratios.
This is also illustrated in  Figure~\ref{abiso}, where we show the abundance difference between the Li \ione\ 6103 \AA\ line and best-fit abundances from Li\ione\ 6707 \AA\ as a function of the isotope ratio.
By requiring the best-fit abundance to be consistent with the $\eps_{Li}$ we obtained from the 6103 \AA\ line, we constrain the $^6$Li/$^7$Li isotope ratio.

The isotope ratio impacts the best-fit abundance significantly, and it also affects the profiles of Li\ione\ 6707 \AA.
There are multiple combinations of $\eps_{Li}$, v$_{\rm mac}$, v$_{\rm r}$ and the corresponding $^6$Li/$^7$Li ratios that provides a reasonable fit of the resonance line (see Fig.~\ref{6707profile}). This degeneracy means that $\eps_{Li}$ and $^6$Li/$^7$Li can hardly be both determined from the resonance line only. We overcome this degeneracy by determining the abundance from the subordinate line. Our method is primarily sensitive to the strength of the Li\ione\ 6707 \AA\ line and it does not rely on the information in the line profile or position. Therefore, it does not require a detailed characterization of the instrument, data reduction, and broadening mechanism.
As a sanity check, we control v$_{\rm mac}$ and v$_{\rm r}$ of our 1D best fit profiles of Li\ione\ 6707 \AA, to make sure that they are consistent with an average v$_{\rm mac}$ = 5.1 $\pm$ 2.5 \kms\ and v$_{\rm r} = -275.1$ \kms\ derived from other spectral lines. 

Our spectral fitting is based on the 1D non-LTE analysis described in Section~\ref{atom}.
Since the best-fit abundances could differ when one uses a 3D model atmosphere, we apply the grid-based correction of \citet{2021MNRAS.500.2159W} on the best-fit abundances as described in Section~\ref{3deffect}.
We note that we focus on the 3D effect on the obtained best-fit abundances and that the line shape difference between 1D and 3D synthetic spectra does not matter in our analysis. 

\begin{table*}
\caption{Non-LTE abundance from the Li\ione\ 6707 \AA\ line as a function of the $^6$Li/$^7$Li isotopic ratio, together with the corresponding best fit macroturbulent and radial velocities.}
\label{tab:67ratio}
\begin{tabular}{lclcr} 
\hline
Reference          &  $^6$Li/$^7$Li, \% &  log A & v$_{\rm mac}$, \kms\ & $\Delta$v$_{\rm r}$, \kms\  \\
\hline                   
test                                                 &  50  & 2.94 & 2.1 & --2.5 \\ 
meteorites, \citet{2003LPI....34.1931M}              &  8.3 & 3.19 & 3.8 & --1.1 \\ 
active K dwarf, \citet{2008ApJ...686..542C}          &  5.0 & 3.27 & 4.2 & --0.8 \\
solar spot, \citet{1997AA...328..695R}               &  3.0 & 3.35 & 4.7 & --0.5 \\ 
test                                                 &  2.0 & 3.41 & 5.0 & --0.4 \\ 
test                                                 &  1.5 & 3.44 & 5.1 & --0.2 \\ 
RGB bump/early-AGB, \citet{2022arXiv220902184K}      &  0.5 & 3.54 & 5.7 & 0.1 \\ 
Spite plateau, \citet{2022MNRAS.509.1521W}           &   0  & 3.64 & 6.1 & 0.2 \\ 
\hline 
\end{tabular} \\
Non-LTE abundance from the Li\ione\ 6103 \AA\ line $\eps$ = 3.44, an average v$_{\rm mac}$ = 5.1 $\pm$ 2.5 \kms\ from other spectral lines.
\end{table*}

\begin{table}
\caption{Impact of the uncertainties on different values on the Li abundance}
\label{tab:lierr}
\begin{tabular}{lllllll} 
\hline
line & \teff & log g & \vt      & continuum & $\Delta_{\rm 3D}$ & total \\
\AA\ &  80 K & 0.13  & 0.2 \kms &  2  \%  &  &  \\
\hline
6103  & 0.05   & 0            & 0                &   0.07      & 0                      & 0.09  \\
6707  & 0.10   & 0            & 0                &   0.04      & 0.02                   & 0.11  \\
\hline
\end{tabular}
\end{table}     

As our measurement of the isotope ratio comes from the best-fit abundances obtained from the two lines, we only need to consider the uncertainties in the abundances, more precisely the uncertainty in the difference between the best-fit abundances obtained from the two lines, and propagate it to the estimate of the isotope ratio.
We adopt 0.1~dex as the uncertainty in the difference between the best-fit abundances taking the followings into account: stellar atmosphere parameters (80~K in \teff, 0.13~dex in log~g, 0.2~\kms\ in \vt), a 0.02~dex uncertainty in the 3D corrections, and a 2 \% shift in the continuum normalisation of the observed spectrum. Uncertainties in \teff\ and continuum placement mostly contribute to the total uncertainty, while changes in log~g and \vt\ produce a negligible ($<$ 0.01~dex) shift in abundance (Table~\ref{tab:lierr}).

Fig.~\ref{abiso} shows the difference between best-fit non-LTE abundances from the  Li\ione\ 6103 and 6707~\AA\ lines as a function of the isotope ratio.
In 1D~non-LTE (blue dotted lines), we find that $^6$Li/$^7$Li = 1.64$^{+1.49}_{-1.08}$ \% provides consistent abundances from the two lines. 
Note that the abundances from the two lines are inconsistent at more than the $2\sigma$ level if we assume $^6$Li/$^7$Li = 0, strongly indicating the presence of $^6$Li in the atmosphere.
The 3D correction increases the best-fit abundance from the Li\ione\ 6707~\AA\ line by 0.17 dex and decrease that from the subordinate line by 0.02 dex, making the detection of $^6$Li more significant and increasing the $^6$Li/$^7$Li ratio to 5.65$^{+5.05}_{-2.51}$ \%.

It is not straightforward to understand the origin of $^6$Li in the investigated Li-rich star.
\citet{1971ApJ...164..111C} mechanism, which is one of the proposed mechanisms for the origin of Li-excess in Li-rich giants, produces only the $^7$Li isotope.
While our 1D non-LTE analysis result does not rule out this scenario, an application of the 3D correction leads to a 0.24~dex difference in the best-fit abundances between the two lines, which is a severe discrepancy given the uncertainties.

Our lithium abundance and isotope ratio determinations are based on the two lines covered by the available observed spectrum. 
However, our analysis would further benefit from observations of other Li\ione\ lines whose strengths are insensitive to the assumed isotope ratio.
Such lines would narrow down the $\eps_{Li}$ range allowed, enabling us to put a stronger constraint on the lithium isotope ratio from the Li\ione\ 6707 \AA\ line.
One such lines is Li\ione\ 8126 \AA, for which we predict  EW = 42 and 37 m\AA\ in non-LTE and LTE, respectively, when using $\eps_{Li}$ = 3.44.

\section{Discussion}
\label{sec:discussion}

The star of interest is likely a single star with mostly normal chemical composition. The exceptions are extremely high Li abundance, an excess of 0.5~dex in [Na/Fe], and potentially low oxygen abundance with an upper limit [O/Fe] $<$ 0.5, which is at least 0.1~dex lower than the typical value. A slightly asymmetric H$_{\alpha}$ profile reveals a signature of disturbance in the stellar atmosphere. The blueshifted core H$_{\alpha}$ may argue for outward moving upper atmospheric layers and mass loss. As for the evolutionary status of the star of interest, its position on the \teff\ -- log~g suggests that the star can be either RGB or AGB star.

AGB stars can exhibit the infrared color excess caused by their mass loss. Some Li-rich giants show the IR excess, and those of them with WISE\citep[Wide-field Infrared Survey Explorer,][]{2013wise.rept....1C} color W1 -- W4 $>$ 1 are the most Li-rich with $\eps_{Li}$ $>$ 2.0 \citep{2015AJ....150..123R}. For the star of interest, the IR color excess W1 -- W4 = 11.093$\pm$0.23 -- {\it 9.113} = 1.98. However, its W4 magnitude should be applied with caution, since it is uncertain and measured with S/N ratio of smaller than 2.

Calculations predict that self-enrichment with lithium through the Cameron-Fowler mechanism may occur at different evolutionary stages: (i) RGB bump stars \citep[for example, ][]{2000A&A...359..563C,2018NatAs...2..790Y}; (ii) upper RGB stars \citep[for example, ][]{2003ApJ...593..509D}; (iii) early AGB stars \citep[for example, ][]{2000A&A...359..563C}. 
Given that the extremely high Li-enhancement is rapidly destroyed and taking into account the results of asteroseismic investigations of Li-rich stars as described in \citet{2021NatAs...5...86Y} where no RGB stars were found with $\eps_{Li}$ $>$ 2.6, we assume that the latter scenario is the most likely for the star of interest. High sodium abundance also supports this guess. \citet{2006A&A...455..291S} investigated a sample of VMP stars and found that the most luminous of their sample giants show higher [Na/Fe] with respect to their fainter counterparts, and they might be AGB but not RGB stars. In those stars, proton capture process converts C and O to N and also Ne to Na, resulting in C and O depletion and Na enhancement. Thus, given the above properties of the star of interest, we conclude that it is likely an AGB star experiencing a Li-flash.

Regarding the potential presence of $^6$Li in the star of interest, it's important to highlight that 2-$\sigma$ detection is achieved when applying 3D corrections based on the calculations from \citet{2021MNRAS.500.2159W}. It is worth noting that these 3D NLTE calculations have never been tested with Li-rich stars. Testing a method is an important step, and employing calculations that have not previously undergone validation with observational data may yield unexpected outcomes. To perform test calculations, one can adopt Li-rich stars where no less than three Li\ione\ lines (6707, 6103, 4602, 8126 \AA, etc.) are detected in their spectra. The criterion of the calculation accuracy is consistent abundances from different Li\ione\ lines in stars with accurate stellar parameters and high quality observed spectra. 

If our detection of $^6$Li is definitively confirmed in future, it may be explained as follows. 
While CF mechanism produces $^7$Li only and operates in stellar interiors, the acceleration of particles in late type stars with chromospheric activity could generate $^6$Li and $^7$Li in the upper photospheric layers \citep{1975ApJ...200..646C,1997SoPh..173..377L}. 
For example, in solar flares, \citet{1997SoPh..173..377L} predicts a temporary enrichment in lithium abundance up to $\eps_{Li}$ = 2, which drops down to typical solar value $\eps_{Li}$ = 2 within three hours. 
Although it is assumed that chromospheric activity decreases with stellar age, \citet{2011PASJ...63S.547T,2016AJ....152...43S} detected He\ione\ 10830 \AA\ line in old stars with [Fe/H] down to $-3.7$. 
The hypothesis of $^6$Li origin in the star of interest via chromospheric activity can be checked by obtaining an additional observed spectrum. 
It could be confirmed or rejected depending on whether variation in Li\ione\ 6707 \AA\ line profile will be found or not.

\section{Conclusions}
\label{sec:con}

We report the discovery of a very metal-poor Li-rich giant star (with effective temperature \teff\ = 4690 $\pm$ 80~K, surface gravity log~g = 1.34 $\pm$ 0.13, metallicity [Fe/H] = $-2.43$ $\pm$ 0.07). We find that its Li abundance $\eps_{Li}$ = 3.42 $\pm$ 0.07 and 3.44 $\pm$ 0.07 in non-LTE~3D and non-LTE~1D, respectively. We construct a model of Li\ione\ atom  based on the accurate  atomic data available to date and perform the non-LTE calculations with this model. To account for 3D effects for Li\ione, we adopt data from \citet{2021MNRAS.500.2159W}.  

From the comparison of the non-LTE abundances from two lines, we determine the isotopic ratio  $^6$Li/$^7$Li = 1.64$^{+1.49}_{-1.08}$ \% in 1D and $^6$Li/$^7$Li to 5.65$^{+5.05}_{-2.51}$ \% when applying the 3D corrections. To our knowledge, this is the first  $^6$Li/$^7$Li measurement in a Li-rich very metal-poor star. 

The proposed method to determine the $^6$Li/$^7$Li isotope ratio relies on the analysis of the resonance Li\ione\ 6707 \AA\ line in conjunction with the subordinate line. Fixing the lithium abundance from the subordinate line that is not sensitive to variations in the $^6$Li/$^7$Li ratio, we overcome a degeneracy between lithium abundance and the $^6$Li/$^7$Li isotopic ratio, which both impact the resonance Li\ione\ 6707 \AA\ line. This method can be applied to other Li-rich stars where no fewer than two Li\ione\ lines can be detected. 

We suggest that the star of interest is likely an early AGB star experiencing a Li-flash. Our interpretation of the lithium enhancement in the star of interest strongly depends on the line formation scenario adopted for the $^6$Li/$^7$Li ratio determination: 1D~non-LTE allows lithium to be produced in the Cameron-Fowler mechanism inside the star, while 3D~non-LTE solidly argues for the presence of a significant amount of $^6$Li, which excludes lithium production in the CF-mechanism. It is worth noting that $^6$Li and $^7$Li can be produced by spallation processes in  atmospheres of stars with chromospheric activity \citep{1975ApJ...200..646C,1997SoPh..173..377L}. However, we postpone the interpretation of the presence of $^6$Li isotope in the star of interest until comprehensive 3D NLTE calculations that account for both isotopes have been verified through testing with Li-rich stars.

In total, we derive abundances for 16 chemical elements from Li to Ba. The investigated star shows high [Na/Fe] = 0.07 $\pm$ 0.03, which is 0.5~dex higher compared to normal stars with similar [Fe/H]. Other chemical element abundances are similar to those found in the literature for very metal-poor stars.

The  star presented here joins the sample of rare Li-rich VMP stars, studies of which can shed light on the mystery of lithium production and its abundance evolution.
The derived abundances and the isotopic ratio can be used as an observational constraint on the poorly known mechanisms of lithium production. For further investigations of the Li-rich stars phenomenon, namely its possible connection with stellar activity together with more robust spectral line formation modeling, observed spectra in a wide wavelength range that covers the He\ione\ 10830 \AA, the Ca\ii\ H and K lines, and the Li\ione\ 8126 \AA\ line would be helpful.

\section*{Acknowledgements}
T.S. acknowledges the Institute of Astronomy, Russian Academy of Sciences, Pyatnitskaya 48, 119017, Moscow, Russia, which made this study possible.
We are indebted to  L.~I. Mashonkina for providing model atoms for the non-LTE calculations and for useful comments on this study. We gratefully acknowledge P.~Bonifacio, Y.~Pakhomov, E.~Ageeva, and B. Nizamov for useful comments and suggestions.
 We are grateful to the reviewer for careful reading the manuscript and for providing valuable comments.
T.S. acknowledges Thomas Nordlander and Ella Wang for clarifying the details of their calculations for Li\ione.
This research is based in part on data collected at the Subaru Telescope, which is operated by the National Astronomical Observatory of Japan. We are honored and grateful for the opportunity of observing the Universe from Maunakea, which has the cultural, historical, and natural significance in Hawaii. 
Z.Y. and N.F.M. acknowledges funding from the Agence Nationale de la Recherche (ANR project ANR-18-CE31-0017) and the European Research Council (ERC) under the European Unions Horizon 2020 research and innovation programme (grant agreement No. 834148).
F.S. thanks the Dr. Margaret "Marmie" Perkins Hess postdoctoral fellowship for funding his work at the University of Victoria.
JIGH acknowledges financial support from the Spanish Ministry of Science and Innovation (MICINN) project PID2020-117493GB-I00.
This work has made use of data from the European Space Agency mission Gaia (https://www.cosmos.esa.int/gaia), processed by the Gaia Data Processing and Analysis Consortium (DPAC, https://www.cosmos.esa.int/web/gaia/dpac/consortium).
This publication makes use of data products from the Wide-field Infrared Survey Explorer, which is a joint project of the University of California, Los Angeles, and the Jet Propulsion Laboratory/California Institute of Technology, funded by the National Aeronautics and Space Administration.

\section*{Author~contribution~statement}
T.S. determined stellar atmosphere parameters, constructed the Li\ione\ model atom, determined chemical composition and led the writing of the manuscript. Z.Y. led the Subaru HR follow-up and T.M. reduced the spectrum. 
\section{Data availability}

The data used in this article will be shared on request to the corresponding authors.


\newpage
Sp.     $\lambda$,\AA\    \eexc, eV   loggf   EW,  m\AA   $\eps_{\rm LTE}$   $\eps_{\rm NLTE}$ \\
 LiI     6103.65   1.85    0.58    71.5    3.27     3.44 \\
 LiI     6707.91   0.00    0.17   426.0    --       3.44 \\
 CH      4313.00   --       --       --    5.56     --   \\
 OI      6300.30   0.00   -9.78     6.6   $<$6.80    $<$6.80 \\
 NaI     5889.95   0.00    0.11   228.4    4.28     3.95 \\
 NaI     5895.92   0.00   -0.19   201.7    4.29     3.91 \\
 MgI     4702.99   4.35   -0.44    67.0    5.15     5.21 \\
 MgI     5528.40   4.35   -0.50    97.9    5.63     5.58 \\
 SiI     4102.94   1.91   -3.14    90.1    5.40     --   \\
 CaI     5588.75   2.53    0.36    83.3    4.37     4.44 \\
 CaI     5857.45   2.93    0.23    29.2    4.00     4.09 \\
 CaI     6102.72   1.88   -0.79    48.9    4.10     4.19 \\
 CaI     6122.22   1.89   -0.31    75.7    4.06     4.12 \\
 CaI     6162.17   1.90   -0.09   104.4    4.34     4.38 \\
 CaI     6439.07   2.53    0.39    81.3    4.22     4.23 \\
 CaI     6493.78   2.52   -0.11    41.0    4.04     4.11 \\
 ScII    4400.38   0.61   -0.54    88.2    0.67     --   \\
 ScII    4415.54   0.60   -0.68    82.1    0.62     --   \\
 ScII    5526.79   1.77   -0.01    47.6    0.67     --   \\
 TiII    4764.52   1.24   -2.69    30.7    2.90     2.94 \\
 TiII    4798.53   1.08   -2.66    49.9    3.02     3.05 \\
 TiII    5013.69   1.58   -2.14    28.9    2.71     2.74 \\
 TiII    5154.07   1.57   -1.75    57.5    2.79     2.79 \\
 TiII    5185.90   1.89   -1.41    49.8    2.72     2.75 \\
 TiII    5336.79   1.58   -1.60    66.9    2.80     2.83 \\
 CrI     5296.69   0.98   -1.36    28.5    2.98     3.40 \\
 MnI     4783.42   2.30    0.03    25.4    2.35     2.69 \\
 FeI     4736.77   3.21   -0.75    53.3    5.02     5.07 \\
 FeI     4871.32   2.87   -0.36    93.4    4.99     5.01 \\
 FeI     4872.14   2.88   -0.57    80.1    4.94     4.98 \\
 FeI     4891.49   2.85   -0.11   105.0    4.97     4.98 \\
 FeI     4903.31   2.88   -0.93    69.3    5.08     5.14 \\
 FeI     4918.99   2.87   -0.34    73.8    4.55     4.57 \\
 FeI     4938.81   2.88   -1.08    55.8    4.96     5.03 \\
 FeI     4966.09   3.33   -0.89    39.0    5.02     5.08 \\
 FeI     5001.86   3.88    0.01    44.0    4.87     4.96 \\
 FeI     5006.12   2.83   -0.63    77.0    4.86     4.90 \\
 FeI     5049.82   2.28   -1.36    79.3    4.96     4.99 \\
 FeI     5068.77   2.94   -1.04    47.2    4.84     4.91 \\
 FeI     5074.75   4.22   -0.20    27.9    5.16     5.25 \\
 FeI     5162.27   4.18    0.02    32.4    4.97     5.07 \\
 FeI     5171.60   1.48   -1.75   106.5    4.97     5.00 \\
 FeI     5191.45   3.04   -0.55    81.3    5.08     5.13 \\
 FeI     5192.34   3.00   -0.52    77.9    4.94     4.99 \\
 FeI     5194.94   1.56   -2.09    87.3    4.96     5.00 \\
 FeI     5215.18   3.27   -0.93    33.1    4.85     4.91 \\
 FeI     5216.27   1.61   -2.10    90.2    5.09     5.13 \\
 FeI     5217.39   3.21   -1.07    21.2    4.66     4.72 \\
 FeI     5232.94   2.94   -0.07   113.2    5.12     5.13 \\
 FeI     5266.55   3.00   -0.39    76.4    4.77     4.80 \\
 FeI     5281.79   3.04   -0.83    55.9    4.88     4.95 \\
 FeI     5283.62   3.24   -0.52    68.6    5.04     5.08 \\
 FeI     5302.30   3.28   -0.88    35.4    4.86     4.91 \\
 FeI     5307.36   1.61   -2.99    39.9    5.02     5.07 \\
 FeI     5324.18   3.21   -0.10    72.5    4.66     4.66 \\
 FeI     5339.93   3.27   -0.68    45.1    4.81     4.86 \\
 FeI     5569.62   3.42   -0.54    49.1    4.90     4.94 \\
 FeI     5572.84   3.40   -0.31    52.7    4.71     4.74 \\
 FeI     5576.09   3.43   -1.00    27.8    4.98     5.04 \\
 FeI     5586.76   3.37   -0.14    71.7    4.84     4.85 \\
 FeI     5615.64   3.33    0.05    84.7    4.85     4.84 \\
 FeI     6024.06   4.55   -0.11    17.4    5.13     5.23 \\
 FeI     6136.61   2.45   -1.50    74.8    5.09     5.12 \\
 FeI     6137.69   2.59   -1.37    63.8    4.94     4.96 \\
 FeI     6173.33   2.22   -2.85    31.2    5.40     5.46 \\
 FeI     6191.56   2.43   -1.42    71.0    4.92     4.94 \\
 FeI     6200.31   2.61   -2.44    19.7    5.20     5.25 \\
 FeI     6213.43   2.22   -2.48    22.3    4.84     4.90 \\
 FeI     6219.28   2.20   -2.44    40.3    5.13     5.19 \\
 FeI     6230.72   2.56   -1.28    78.6    5.07     5.08 \\
 FeI     6252.55   2.40   -1.76    65.8    5.13     5.17 \\
 FeI     6265.13   2.18   -2.55    31.8    5.05     5.11 \\
 FeI     6297.79   2.22   -2.74    23.3    5.11     5.17 \\
 FeI     6301.50   3.65   -0.72    28.0    4.93     5.00 \\
 FeI     6335.33   2.20   -2.23    49.1    5.05     5.11 \\
 FeI     6336.82   3.69   -0.86    18.1    4.86     4.93 \\
 FeI     6393.60   2.43   -1.43    62.2    4.75     4.80 \\
 FeI     6400.00   3.60   -0.52    41.7    4.92     4.98 \\
 FeI     6408.02   3.69   -1.00    14.4    4.88     4.95 \\
 FeI     6421.35   2.28   -2.01    54.6    5.02     5.06 \\
 FeI     6430.84   2.18   -1.95    61.4    4.94     5.00 \\
 FeI     6481.87   2.28   -2.98    22.1    5.38     5.44 \\
 FeI     6494.98   2.40   -1.27    93.8    5.12     5.16 \\
 FeI     6593.87   2.43   -2.39    21.2    4.95     5.01 \\
 FeII    5197.57   3.23   -2.24    42.1    4.92     4.92 \\
 FeII    5234.62   3.22   -2.17    59.2    5.14     5.14 \\
 FeII    5276.00   3.20   -2.10    60.6    5.06     5.06 \\
 FeII    5284.10   2.89   -3.09    27.7    5.07     5.07 \\
 FeII    5534.84   3.24   -2.75    21.8    5.00     5.00 \\
 FeII    6516.08   2.89   -3.32    17.4    4.97     4.97 \\
 NiI     5035.36   3.64    0.29    29.0    3.69     --   \\
 NiI     6643.63   1.68   -2.22    28.4    3.71     --   \\
 ZnI     4810.53   4.08   -0.14    29.7    2.29     2.44 \\
 SrII    4077.72   0.00    0.15   194.2    0.36     0.39 \\
 SrII    4215.52   0.00   -0.14   168.3    0.35     0.36 \\
 YII     4883.68   1.08    0.07    34.3   -0.54    -0.40 \\
 YII     5205.73   1.03   -0.34    22.1   -0.51    -0.37 \\
 BaII    5853.67   0.60   -1.00    28.4   -0.84    -0.83 \\
 BaII    6141.72   0.70   -0.08    67.9   -1.00    -0.98 \\
 BaII    6496.90   0.60   -0.38    78.5   -0.65    -0.61 \\
 BaII    4554.03   0.00    0.17   123.7   -1.01    -1.07 \\

\end{document}